\pdfoutput=1
\documentclass[sigconf,authorversion,nonacm]{acmart} 
\AtBeginDocument{%
  \providecommand\BibTeX{{%
    \normalfont B\kern-0.5em{\scshape i\kern-0.25em b}\kern-0.8em\TeX}}}

\usepackage{romannum} 
\AtBeginDocument{\pagenumbering{arabic}} 
\usepackage{xspace}
\newcommand{\tool}[0]{\textsc{BotDesigner}\xspace}
\newcommand{\method}[0]{\textit{Conversation Regression Testing}\xspace}

\renewcommand{\paragraph}[1]{\vspace{.6em}\noindent\textbf{#1}\hspace*{.5em}}
\usepackage{enumitem}
\usepackage{booktabs}
\usepackage{xcolor}
\usepackage{multirow}
\usepackage{array}
\usepackage{siunitx}
\usepackage{graphicx}

\usepackage{dblfloatfix}

\begin{document}

\title[Conversation Regression Testing]{Conversation Regression Testing: \\A Design Technique for Prototyping Generalizable Prompt Strategies for Pre-trained Language Models}

\author{J.D. Zamfirescu-Pereira}
\affiliation{%
\institution{UC Berkeley}
\city{Berkeley}
\state{CA}
\country{USA}
}
\email{zamfi@berkeley.edu}

\author{Bjoern Hartmann}
\affiliation{%
\institution{UC Berkeley}
\city{Berkeley}
\state{CA}
\country{USA}
}
\email{bjoern@eecs.berkeley.edu}

\author{Qian Yang}
\affiliation{%
\institution{Cornell University}
\city{Ithaca}
\state{NY}
\country{USA}
}
\email{qianyang@cornell.ed}


\begin{abstract}
Pre-trained language models (LMs) such as GPT-3 can carry fluent, multi-turn conversations out-of-the-box, making them attractive materials for chatbot design. Further, designers can improve LM chatbot utterances by prepending textual \textit{prompts} -- instructions and examples of desired interactions -- to its inputs.
However, prompt-based improvements can be brittle; designers face challenges systematically understanding how a prompt strategy might impact the unfolding of subsequent conversations across users.
To address this challenge, we introduce the concept of Conversation Regression Testing. Based on sample conversations with a baseline chatbot, Conversation Regression Testing tracks how conversational errors persist or are resolved by applying different prompt strategies. 
We embody this technique in an interactive design tool, \textit{BotDesigner}, that lets designers identify archetypal errors across multiple conversations; shows common threads of conversation using a graph visualization; and highlights the effects of prompt changes across bot design iterations.
A pilot evaluation demonstrates the usefulness of both the concept of regression testing and the functionalities of BotDesigner for chatbot designers.
\end{abstract}




\begin{teaserfigure}
  \centering
  \includegraphics[width=\textwidth]{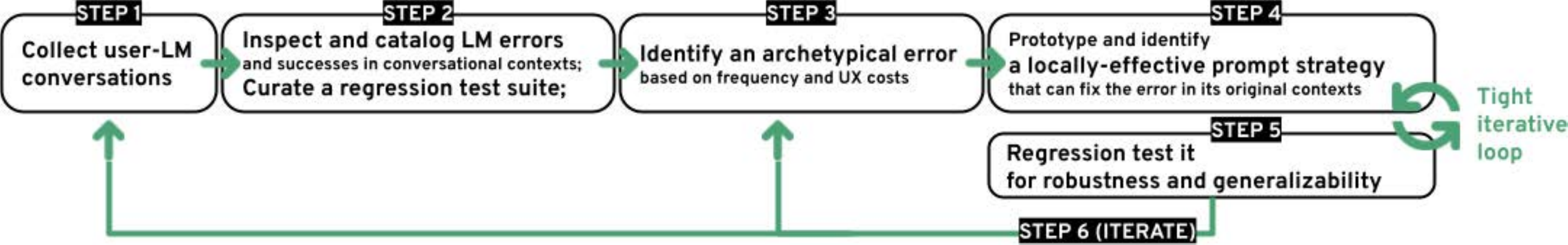}
  \caption{We propose \method, a workflow for chatbot designers to systematically experiment and evaluate various prompt strategies' impact on pre-trained-language-model-powered conversational interactions. We also present \tool, a prompt prototyping tool that operationalizes this workflow.}
  \vspace{0.2cm}
  \label{fig:teaser}
\end{teaserfigure}

\maketitle

\section{Introduction}

The combination of pre-trained large language models (LM) and prompts offers exciting new opportunities for chatbot design.
Recent pre-trained LMs (\text{GPT-3}~\cite{GPT3-LMs-are-FSL}, \text{GPT-J}~\cite{gpt-j}, \text{Jurassic-1} \cite{lieber2021jurassic}, and \text{T0}~\cite{T0-Sanh2021}) can engage in fluent, multi-turn conversations out-of-the-box.
Removing the costs and data requirements for training supervised models, these models substantially lower the barrier of entry for creating a passable conversational user experience (UX)~\cite{Stanford_foundation_models_white_paper}.
Further, chatbot designers can improve LM outputs by prepending \textit{prompts}---textual instructions and examples of designers' desired interactions---to LM inputs (Table \ref{tab:prompt-example}.) Prompts directly bias the model towards generating the desired outputs, raising the ceiling of what conversational UX is achievable with little or no labeled data.
In the past two years, the promises of this new \textit{pretrain-and-prompt} paradigm have been propelling a rapidly growing set of prompt design tools~\cite{promptsource,logan2021cutting,gehman-etal-2020-realtoxicityprompts}, research across HCI/UIST (e.g.,~\cite{GenLine-UIST21,AIchain-CHI22,AIchain-CHI22lbw,ChilltonCHI22-prompt-guideline-txt-img,coauthor-chi22}) and NLP (e.g.,~\cite{instructGPT,liu2021pretrain,T0-Sanh2021,schick2021true,lang2022cotraining}), and even end-user-facing applications~\cite{GenLine-UIST21}.
Some scholars even argued that this paradigm represents a ``\textit{sea change}'' in NLP and will ``\textit{replace}'' the fully-supervised and pre-train-and-fine-tune paradigms~\cite{liu2021pretrain}.

\begin{table}[ht]
\centering
\renewcommand{\arraystretch}{1.2} 
\begin{tabular}{p{0.3\linewidth} p{0.6\linewidth}} 
\toprule
\textbf{Prompt Strategy} & \textbf{Resulting Conversation}

  \\ 
\hline
No prompt\newline (baseline) & 
\includegraphics[height=12px]{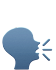} Ok hang on while I get a chair\newline
\includegraphics[height=12px]{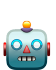} Scoot to the front of your chair\textit{[...]}
\\ 
 \hline
Explicit instruction\newline to wait  &
\textcolor{blue}{\includegraphics[height=12px]{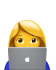} If the user asks you to wait, explain that this is not a problem \textit{[...]}} \newline
\includegraphics[height=12px]{fig/personji.pdf} Ok hang on while I get a chair\newline
\includegraphics[height=12px]{fig/robotji.pdf}\colorbox{green!30}{Once you have your chair,} scoot to the front of it\textit{[...]}
\\
\bottomrule
\end{tabular}
\captionsetup{width=0.95\linewidth}
\vspace{0.2cm}
    \caption{An example of how designers can directly improve chatbot interactions by modifying \textcolor{blue}{prompt strategies} (\includegraphics[height=10px]{fig/designerji.pdf}). Note the change in the bot's (\includegraphics[height=10px]{fig/robotji.pdf}) response to the user's (\includegraphics[height=10px]{fig/personji.pdf}) statement.}
    \label{tab:prompt-example}

\end{table}

The pre-train-and-prompt paradigm also brings a series of new challenges to chatbot design, namely, the challenges of \textit{designing robust prompt strategies.}
To leverage prompts for production-level chatbots, designers need to first systematically assess (1) in what conversational contexts the pre-trained LM is likely to fail and (2) how frequent or damaging each failure or failure mode is, in order to identify the right problems to solve with prompts. 
Next, designers need to (3) identify a prompt strategy that can fix the target failure in its original conversational context, and finally, to assess its \textit{generalizability} and \textit{robustness} systematically, that is, assessing (4) whether it can fix similar failures in other conversational contexts, and whether it might cause new errors across the numerous ways the conversations can unfold subsequently for different users. These are challenging tasks~\cite{T0-Sanh2021,liu2021pretrain,Stanford_foundation_models_white_paper}.

A few HCI researchers have started to create workflows and tools that aid prompt strategy design, for example, for human-LM collaborative writing \cite{AIchain-CHI22lbw}.
However, such workflows and tools for chatbots are extremely rare. 
Instead, chatbot designers often experimented prompts \textit{ad-hoc} using tools such as GPT-3 Playground~\cite{GPT3playground}; Some even treated prompt strategy design as ``\textit{rolling the dice}''~\cite{AIchain-CHI22}. 
It remains unclear how designers can holistically analyze the highly-contextual errors LMs make across conversations (challenges 1, 2), or how they can resolve the errors without unknowingly causing new errors in preceding or subsequent conversations (challenges 3, 4).

As a step toward more systematic and rigorous prompt strategy prototyping, we introduce the concept of \method. 
Taking inspiration from software regression testing, \method uses the conversational contexts where a baseline LM has failed (or notably succeeded) as reusable test cases and helps designers track the effects of prompt strategy updates on these test cases.
This approach allows designers to freely experiment with many prompt strategies to address a particular error in context, while ensuring the system's overall stability and a trajectory of continuous improvements. 

Operationalizing this concept, we then present \tool, a prompt strategy prototyping tool that integrates the \method workflow into an  interactive machine learning analysis tool (one that tracks model performance across iterations and provides insights into what changes yield what performance improvements.) Such tools have shown remarkable traction with designers and developers in non-conversational domains (e.g., Weights and Biases~\cite{wandb}).
\tool consists of four components:

\begin{itemize}[leftmargin=*]
\item \textbf{Conversation Collector}, an interface for collecting sample conversations between a baseline LM-based chatbot and real-world users (or crowd workers);
\item \textbf{Annotator}, an interface for inspecting and cataloging the problematic (or successful) utterances made by the baseline bot, across many conversations with multiple users. These errors are opportunities for prompts to help, as well as test cases for \method;

\item \textbf{Visualizer}, a graphic visualization that aids designers to identify archetypal errors by showing the baseline bot's failures and successes against the backdrop of common end-user-LM conversation patterns. These archetypal errors help designers to prioritize their prompt design efforts;
\item \textbf{Regression Tester}, features that embody \method. When designers experiment with a new prompt strategy, these features enable them to track whether the target error persists or gets resolved, or if new errors have appeared, as a result of the new strategy.
\end{itemize}

This paper presents the concept of \method, the implementation of \tool, and a small user evaluation study that preliminarily demonstrates the usefulness of both for chatbot designers when designing instructional chatbots.

This paper makes two contributions, one conceptual and one technical. 
The primary contribution is the concept of \method for prompt strategy design. While most prior work focused on exploration-and-ad-hoc-testing stage of prompt design, \method offers an initial workflow to for assessing prompt strategies' robustness and generalizability.
Secondly, the technical contribution of this paper lies in the techniques for implementing \tool. It presents a novel conversation visualization technique that visualizes common conversation patterns across many discrete conversations between an LM and various users. It can be useful for developing many other human-LM interaction analysis or design tools.
\tool also implements an interface for \method, a technique that can be valuable for prototyping prompts for many other LM applications beyond conversational interactions.

\section{Related Work}

We briefly review three threads of related work: 1) workflows and tools for interactively improving NLP model performance and 2) for improving conversational UX, and finally 3) prior conversation visualization techniques and analytical tools.

\subsection{NLP Modeling Workflows and Tools}

NLP modeling workflows and tools roughly fall under three categories~\cite{liu2021pretrain}.
\textit{\textbf{Fully supervised learning}}, where a task-specific model is trained on a dataset of input-output examples for the task, has long played a central role in machine learning (ML) and natural language processing (NLP).
Because fully labeled datasets are often insufficient for learning high-quality models, interactive NLP tools for improving model performance focused heavily on assisting feature engineering; providing models with the appropriate inductive bias to learn from this limited data.
Towards this goal, supervised NLP tools most often embodied one of the two workflows:
\begin{itemize}[leftmargin=*]
    \item Tools such as LightSIDE \cite{mayfield2013lightside} assist NLP modelers to define and extract salient features from raw data. These tools adopted a five-step workflow that many seminal interactive ML tools (e.g., Crayons~\cite{fails2003interactive}, ModelTracker~\cite{amershi2015modeltracker}, Gestalt~\cite{gestalt}, and Weights and Biases~\cite{weights-biases}) have pioneered: Modelers (\romannum{1}) inspect raw data; (\romannum{2}) label data or extract features from the data, sometimes with the assist of ML; (\romannum{3}) train an initial model, (\romannum{4}) classify, view, and correct the model's outputs, and (\romannum{5}) iterate on this process while the tools track the model's performance improvements and provide insight into what changes yield the improvements.
    \item The second workflow emerged in response to the criticism that the above workflow left out considerations of ML amateurs\cite{powertothepeople}. 
    Researchers created ``\textit{human-centered ML tools}'' that added end-users to every step of the first workflow (e.g., allowing them to provide traces of their natural interaction with the model for model training~\cite{adaptiveUI} and transfer learning~\cite{mishra-CHI2021-transfer}, nominate features \cite{flock}, demonstrate desired model behaviors~\cite{machine-teaching-yang-DIS18}, etc.)
    These tools demonstrated that integrating an understanding and natural interaction data of end-users into ML workflow can improve both UX and model performance~\cite{powertothepeople}.\label{enduserML}
\end{itemize}

In 2017-2019, the standard way of NLP modeling shifted to ``\textit{\textbf{pre-train and fine-tune}}'', with fully supervised learning playing an ever-shrinking role~\cite{liu2021pretrain}.
This paradigm embodies a two-step-only, no-longer-task-specific ML workflow.
\begin{enumerate}
    \item[Step \romannum{1}] Modelers pre-train a model with a fixed architecture on large, unlabeled textual data. In this process, the pre-trained LM learns general-purpose language features that can be used for a wide range of tasks (e.g., predicting the next line of code or prose, document summarization, biomedical question answering, translation, and more.)
    GPT \cite{GPT3-LMs-are-FSL,gpt-j} and BERT \cite{BERT} exemplify families of pre-trained LMs.
    \item[Step \romannum{2}] Modelers adapt the pre-trained LM to the particular interaction task at hand through fine-tuning.
\end{enumerate}

In this paradigm, the main focus of model tuning turned from feature to objective engineering, designing the training objectives for both pre-training and fine-tuning.
As a result, most aforementioned interactive ML tools no longer apply.
While a few commercial general-purpose ML tools (e.g., Azure~\cite{Azure-for-finetune-BERT}) can support this new workflow, we did not find interactive NLP tools tailored for this workflow in our literature search. 

The past two years have been witnessing another paradigm shift in NLP: the rise of the ``\textbf{\textit{pre-train, prompt, and predict}}'' paradigm~\cite{liu2021pretrain}.
This paradigm follows roughly the 2-step workflow above. 
However, instead of adapting pre-trained LMs to particular tasks via objective engineering, modelers reformulate the tasks to look more like those solved during the original LM training with the help of a textual \textit{prompt}. 
For example, GPT-3 can automatically translate users' natural language requests to html code using the prompt template/strategy
``\texttt{web code description: <natural language request> html:<html> css: <css> javascript: <js>}''~\cite{GenLine-UIST21}.
Modelers curate a large set of such prompts using a template and retrain the LM with them~\cite{openprompt,promptsource}.

In this paradigm, the main focus of model tuning turned to prompt engineering, designing the appropriate prompts and prompt strategies that yield the desired model behaviors.
Further, because many prompts are human-readable, prompts also present renewed opportunities to engage end-users in the modeling process. Tools have emerged to enable crowd workers or end-users to contribute queries and prompt strategies \cite{promptsource, openprompt}. 

Noteworthily, even for experts, identifying robust and generalizable prompt strategies requires extensive trial and error, where modelers iteratively experiment and assess the effects of various prompt strategies on concrete input-output pairs, before assessing them more systematically on large conversation datasets.
A well-established prompt design workflow does not yet exist. How a prompt or a prompt strategy may directly impact model outputs, or how it modifies pre-trained LM's billions of parameters during re-training, are both active areas of NLP research~\cite{T0-Sanh2021,liu2021pretrain}.

\subsection{Prototyping Chatbot UX}\label{current-UX}
A well-established workflow exists for designing and prototyping multi-turn conversational interactions and experiences (``\textit{chatbot UX}'', for short)~\cite{pricilla2018designing,chatbot-migrants-DIS20,woz-living-room-speech,Suede-woz-speech,storytellingwizard, cranshaw2017calendar}). 
Following this workflow, chatbot designers first (\romannum{1}) identify the chatbot's functionality or persona and draft ideal user-bot conversations, for example, through Wizard-of-Oz or having experts drafting scripts; (\romannum{2}) create a dialogue flow template (e.g., ``\textit{greeting message, questions to collect user intention, ...}''); and finally (\romannum{3}) fill the template with supervised NLP models (e.g., user intention classifier, response generator, etc.)
Many tools that support this process exist supporting this process, for example, Google Dialogflow and Facebook Messenger tools for step (\romannum{2}) and (\romannum{3}).

While highly valuable, these conversation-template-oriented tools are ill-fitted for pre-trained LMs.
However, chatbot design tools for the pre-train-and-prompt paradigm are extremely rare.
The closest related work is \textit{AI Chains}\cite{AIchain-CHI22}, a tool for exploring human-LM collaborative writing interactions.
It allows designers to construct a chain of LMs where the output of one LM becomes the input for the next, and to test the resulting interactions themselves. The tool successfully enabled designers to explore prompt and chaining strategies more efficiently and strategically~\cite{AIchain-CHI22lbw}. However, it is unclear whether the resulting strategies were effective or robust beyond the few interactions contexts that the designers experimented with.

\subsection{Conversation Visualization and Analysis}

Prior work on visualizing conversations has either focused on visualizing the structure of a dyadic (email)~\cite{venolia2003understanding} or multi-party conversation~\cite{donath-visualizing-conversations-2002, zhao2012no} over time (newsgroups, etc); or, they've sought to create a more abstract, higher-level picture of the topics covered in a conversation~\cite{bergstrom-conversation-clusters-2009}. Our needs here are different, since we're considering the unique settings of multiple independent conversations about the same topic---visualizing which pieces are shared and which are unique to each conversation. Some related work does also touch on the adjacent task of visualizing the structure of multiple \textit{tutorials} (rather than conversations) covering a single topic, exploring which pieces are shared and which are unique to each tutorial~\cite{kong2012delta,pavel2013browsing}.

\section{Conversation Regression Testing}

We wanted to help chatbot designers to freely prototype and systematically evaluate prompt strategies, thereby empowering them to leverage pre-trained LMs and prompts in their design.
To this end, we introduce the concept of \method. 

\subsection{Definition and Benefits} \label{sec:crt-def}

\begin{figure*}[t]
    \centering
    \includegraphics[width=\textwidth]{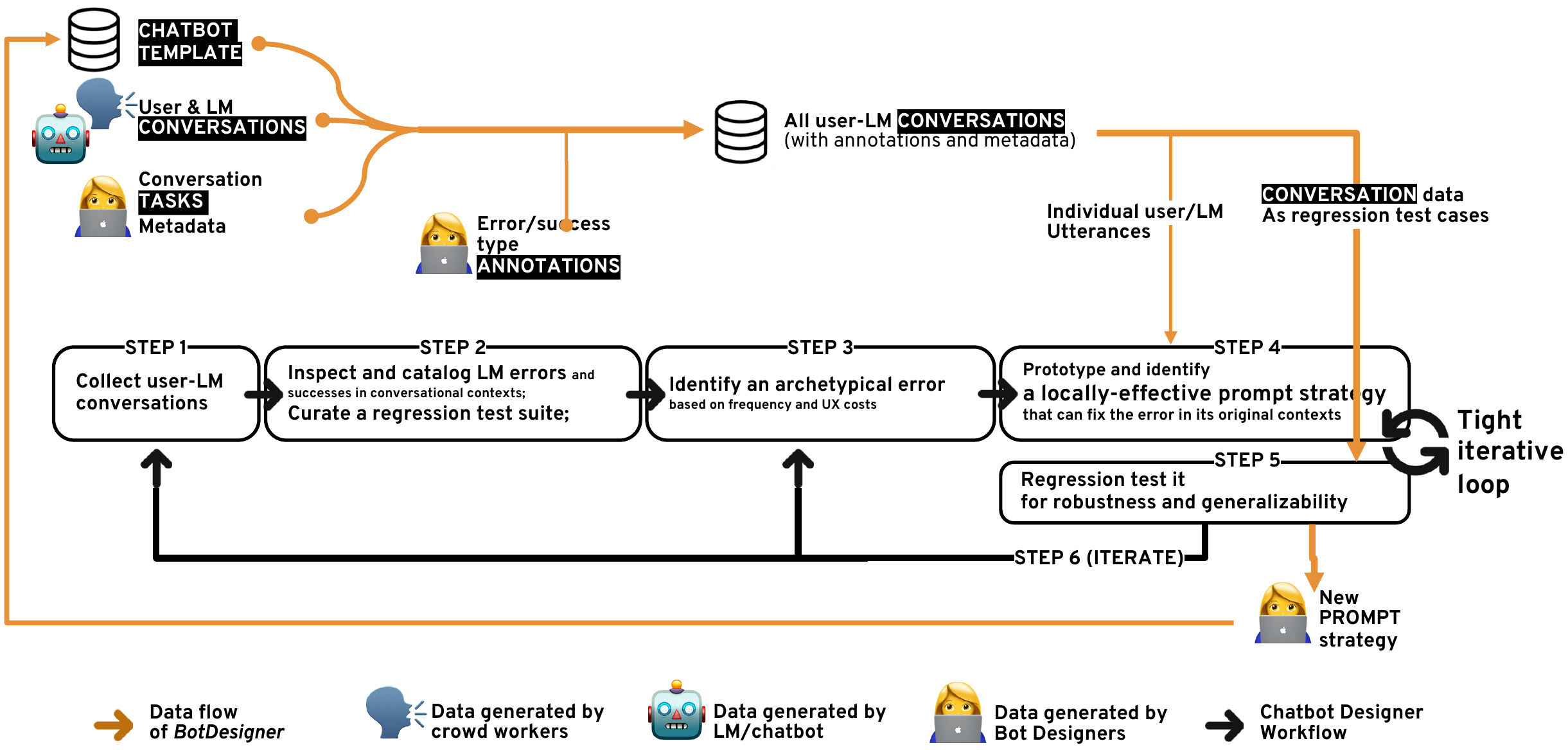}
    \vspace{0.2cm}
    \caption{\method workflow and \tool data flow.}
    \label{fig:crt-hero-figure}
\end{figure*}

\method is an iterative workflow for prototyping and evaluating prompt strategies.
Following this workflow, chatbot designers start by identifying a baseline prompt strategy (or an off-the-shelf pre-trained LM, i.e. with no prompt strategy). They then carry out the following complementary activities (Figure~\ref{fig:crt-hero-figure}):

\begin{enumerate}[leftmargin=*]

    \item \textit{Collect human-LM conversations:} Collect a diverse set of conversations between the baseline LM and end-users through crowdsourcing or in-person user studies;
    
    \item \textit{Inspect and catalog LM errors and successes in context:} Inspect the errors and successes both in the contexts where they occurred and in aggregate, across the myriad ways the baseline user-LM conversations have unfolded; add noteworthy user-LM conversation turns to a suite of regression test cases;
    
    \item \textit{Identify an archetypical error} based on how frequent or damaging each error or error pattern is; develop intuitions of possible new prompt strategies for addressing the error;

    \item \textit{Identify a locally-effective prompt strategy:}~Experiment with new prompt strategies to fix a particular archetypical error in the conversational context where it originally occurred; Identify one locally effective prompt strategy;

    \item \textit{Regression test for robustness and generalizability:}~Apply the locally effective prompt strategy to the entire regression test suite, inspecting its robustness (whether it has fixed similar failures in other conversational contexts) and generalizability (whether it has caused new errors across the numerous ways the conversations can unfold subsequently for different users). If not, iterate on step 4-5 or even collect more conversations (step 1-5) before proceeding. If positive for both, continue;
    
    \item \textit{Iterate while tracking:}~Consider the robust and generalizable prompt strategy as a new baseline, iterate on the whole process (step 1-6) while tracking which errors have been resolved versus persisted.

\end{enumerate}

Central to this workflow are the concepts of \textit{conversation regression testing} and \textit{prompt prototyping in human-LM conversational contexts}.
They highlight the benefits of \method over existing common practices.

\paragraph{Benefits over current chatbot UX prototyping workflow.~}
Similar to software regression test suites~\cite{brooks1995mythical}, conversation regression test suites enable chatbot designers to track the effects of prompt strategy updates on many discrete conversations with different users. 
This approach is particular valuable for prompt strategy design, because UX improvements and breakdowns caused by prompts are often brittle. 
In comparison to the current UX practice where designers tend to test their prompt strategies on the utterances they themselves authored in an ad-hoc manner \cite{AIchain-CHI22lbw,ChilltonCHI22-prompt-guideline-txt-img}, conversation regression test cases enable designers to freely experiment with many prompt strategies, without unknowingly causing new errors in preceding or subsequent conversations.

Importantly, \method is not \textit{merely} Regression Testing applied to prompt design. \method is a rapid and iterative prototyping process. Each iteration resolves an error or an error mode without regression. This is 
different from software regression tests, whose use is typically limited to when new program updates reintroduces \textit{old} errors (hence the name \textit{regression}.)

\paragraph{Benefits over current NLP practice.~} 
\method highlights the importance of the use of user-LM conversation texts throughout the prompt strategy prototyping process. Designers inspect errors and test new strategies, both in the original user-LM conversational contexts where errors (or successes) occurred.
This is a departure from current common NLP practice, where modelers typically evaluated prompt strategies on pre-curated human-human conversation datasets. Taking a lesson from human-centered ML work, end-user interactions with a model -- particularly their reactions to its errors -- should not be an afterthought.

\subsection{\method In Practice: An Example Design Process} \label{sec:example}

\begin{table}[b]
    \renewcommand{\arraystretch}{1.2}
    \resizebox{0.95\linewidth}{!}{
        \begin{tabular}{p{\linewidth}}
        \toprule
         \textbf{Baseline prompt}\\
         \hline
            \texttt{Consider the following set of exercises:}
            
            \texttt{1. Tricep Dips. Scoot to the front of your chair, with both hands facing forward, [...]}
            
            \texttt{2. Seated Leg Lifts. Grab the sides of your chair [...]}
            
            \texttt{[...]}
            
            \texttt{Instruct the user in completing each exercise step-by-step.}
            
         \\ \hline
         \textbf{New prompt (fixing the ``skip a step error'')}
         \\ \hline
         \texttt{Consider the following set of exercises:}
            
            \texttt{1. Tricep Dips. Scoot to the front of your chair, with both hands facing forward, [...]}
            
            \texttt{2. Seated Leg Lifts. Grab the sides of your chair [...]}
            
            \texttt{[...]}
            
            \texttt{Instruct the user in completing each exercise step-by-step.}
            
            \textcolor{blue}{\texttt{Don't skip any steps.}}
         \\ \hline
         
        \end{tabular}
    }
    \vspace{0.2cm}
    \caption{The baseline and improved prompts in the \textit{ExerciseBot} design example. }
    \label{tab:example-baseline-bot}
\end{table}

Let us ground the concepts and workflow of \method and their benefits in a concrete example.
Consider ourselves chatbot designers who are creating an \textit{ExerciseBot}, a voice-based conversational agent that walks users through a set of physical exercises that they can perform at their desk.
Following the \method workflow, we can rapidly prototype various prompt strategies in-context and systematically evaluate their robustness and generalizability:

We start by identifying a baseline prompt strategy. Here we use GPT-3's \texttt{text-davinci-001} model (setting \textsc{temperature} $= 0$) out-of-the-box.
We use the simple combination of a set of publicly available exercise instructions and a request to ``\textit{instruct the user in completing each exercise step-by-step}'' as our baseline (Table \ref{tab:example-baseline-bot}); 

\begin{enumerate}[leftmargin=*]

    \item \textit{Collect human-LM conversations:} We collect 30 conversations between the baseline bot and 10 Mechanical Turk workers, which yields many creative yet realistic utterances that we could hardly anticipate (``\textit{At my age I'm going to have to break them up.}'' ``\textit{Is it more effective to do all [exercises] at once?''}).

    \item \textit{Inspect and catalog errors and successes in context:} We found that the baseline prompt strategy is sufficient to create a passable chatbot that, most often, naturally walked users through the exercise step-by-step (e.g., User: ``\textit{At my age I'm going to have to break them up.}'' Bot: ``\textit{That's ok, just try to complete all 5 reps.}'')
    
    We also identified a number of error patterns. For example, the ``skip a step'' error is that the bot skips a step when walking users through the exercises.

    The ``\textit{unsympathetic}'' error is where the bot routinely ignores user requests for help (``\textit{Can we try an easier exercise?}'') or expressions of distress (``\textit{Ow, that hurt!}''.) We collected these conversations as substrates for our \method test suite.

    \item \textit{Identify an archetypical error.~}  We chose to focus on the ``\textit{skip a step}'' error, since it causes confusion if not physical danger during the exercises. It also has frequently caused breakdowns in subsequent conversations when users requested clarifications. 

    \item \textit{Identify a locally-effective prompt strategy:}~After extensive experimentation, we resolved the ``\textit{skip a step}'' error by simply appending the explicit instruction ``\texttt{Don't skip any steps.}'' to the end of the baseline prompt, before the user-bot conversations begin. Another locally-effective strategy is to number the sub-steps within each step of the exercises in the initial prompt (Table \ref{tab:example-baseline-bot}.)

    \item \textit{Regression test for robustness and generalizability:}~Applying the two new strategies to the previously curated test cases, we noticed that the explicit instruction strategy consistent resolves the ``\textit{skip a step}'' error, while the numbering-the-steps strategy only worked for some exercises. However, in some contexts, the explicit instruction strategy caused a side effect: It makes the bot's stubbornly stick to the step-by-step exercise instructions, even when users said this step is too hard. It could worsen the ``\textit{unsympathetic}'' error.

    With this trade-off in mind, we iterate on step 4-5, exploring additional prompt strategies that may work even better.

    We could also choose to collect additional conversations (for example, on a different set of exercises), thereby identifying new patterns of errors and success (steps 1 and 2). This approach allows us to fully understand the extent to which the new prompt strategy is robust and generalizable before adopting it.
    
    \item \textit{Iterate on this process to tackle additional errors} while tracking ExerciseBot's behavior changes using the \method test suite.

\end{enumerate}

\section{\tool: A Tool that Operationalizes \method}

We present \tool, a chatbot prompt strategy prototyping tool that operationalizes the \method workflow described in \S\ref{sec:example}.

\subsection{System Overview}

\tool enables \method with the following functionality:

(1) A \textbf{conversation collection} interface that enables the crowdsourcing of a set of baseline conversations with a baseline GPT-3 based chatbot; this interface enables step (1) described in~\S\ref{sec:example}.

(2) A \textbf{conversation visualization} and \textbf{annotation} interface that shows conversation flow across multiple users' conversations (for a single \textit{task}, defined in \S\ref{sec:inputs}) using a graph interface, highlighting which utterances are common across conversations, and aiding in the categorization and tagging (annotation) of individual problematic or particular successful bot-provided utterances for targeted improvement or maintenance. This interface enables steps (2)-(3) from~\S\ref{sec:example}.

(3) A \textbf{utterance testing} interface that situates individual problematic utterances in context and highlights changes to those utterances caused by updates to the bot. This interface enables steps (4)-(5) from~\S\ref{sec:example}.

In conjunction with a built-in code editor, these interfaces support iteration over chatbot prompt designs. 

\subsection{Inputs} \label{sec:inputs}

\tool relies on three types of input data: \textbf{conversations}, \textbf{tasks}, and \textbf{templates}, representing, respectively, individual multi-turn user interactions with a specific bot (\textit{conversation}), a set of structured instructions that make up the user's task (\textit{task}), and a set of prompts comprising a specific point design for a chatbot (\textit{chatbot template}).
Although we believe \method can be usefully applied to any type of chatbot, we chose to focus on \textit{task-oriented instructional} interactions because of the opportunities for aggregation offered by similarities across multiple conversations by multiple users focused on the same task. 

\textbf{Conversations} are specific multi-turn interactions collected by \tool, consisting of a dialog data structure that includes each conversation partner's utterances as well as any error annotations provided \textit{post facto} by the designer or human conversation partner. Each conversation is attached to the specific template and recipe used to generate the bot's utterances.

\textbf{Tasks} are specific structured task descriptions comprised of a name, description, and set of steps the user is expected to complete. Some tasks may also include metadata such as a list of the items required to complete the task.

\begin{figure}
    \centering
    \includegraphics[width=\columnwidth]{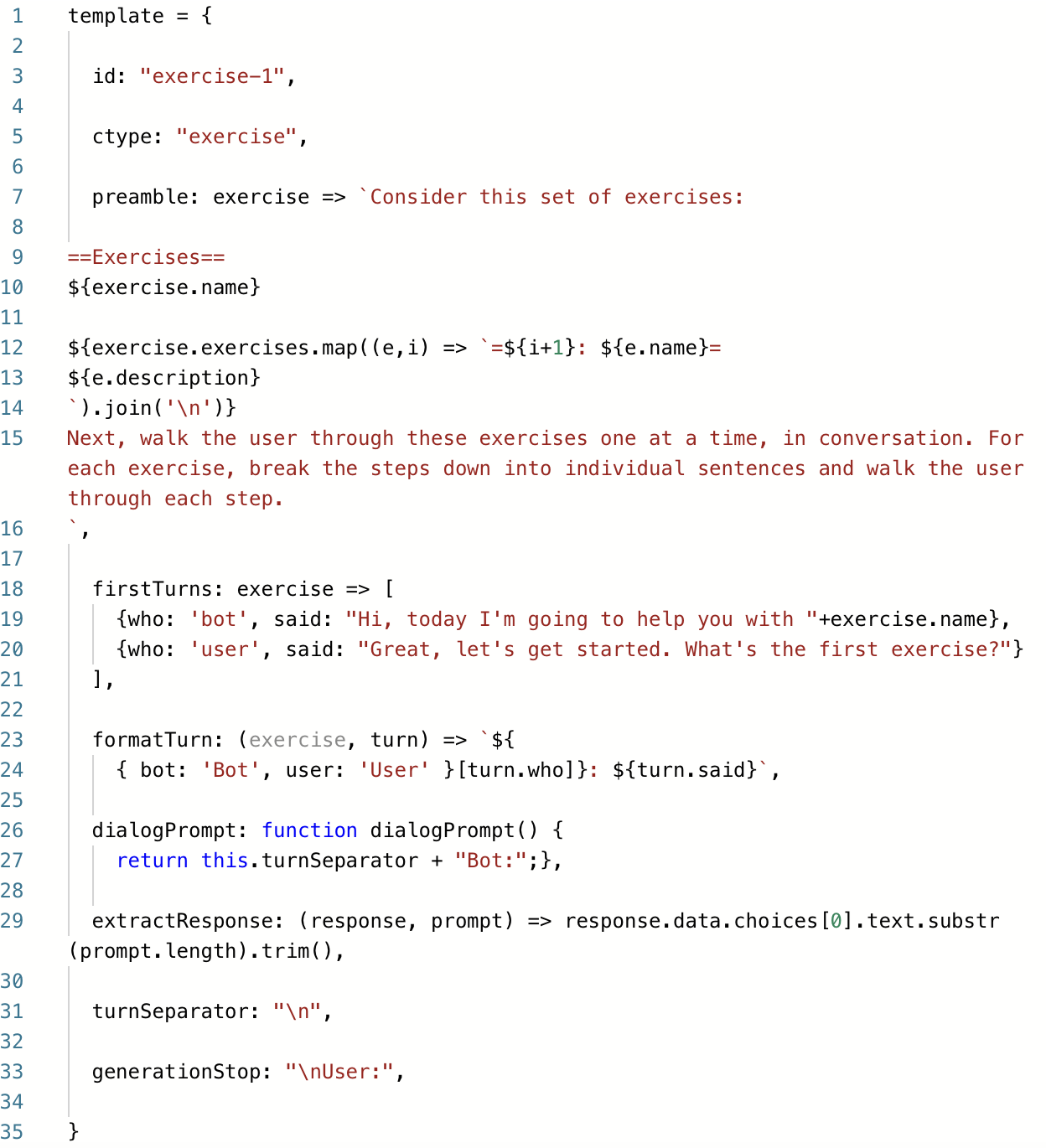}
    \caption{The various properties of this chatbot template describe the prompt \textit{preamble} (the text prepended to the conversation dialog), instructions for \textit{formatting} prior conversational turns into the LM prompt, instructions for \textit{prompting} and \textit{extracting} the chatbot utterance from the LM's prediction, and other assorted parameters.}
    \label{fig:template}
\end{figure}

\textbf{Chatbot templates} describe the set of prompts that are sent as a prefix to the backing LM (GPT-3 in the case described here). Each template contains instructions for (1) how to convert a structured \textbf{task} of the appropriate type into plain text, suitable for inclusion into the LM text prompt, and (2) code describing how to lay out, in the prompted text, the turn-by-turn dialog-in-progress that is stored in the \textbf{conversation}. Templates also describe how the LM output should be parsed and the bot's response utterance extracted. See Fig.~\ref{fig:template} for an example.

\subsection{Using \tool}

\begin{figure}
    \centering
    \includegraphics[width=\columnwidth]{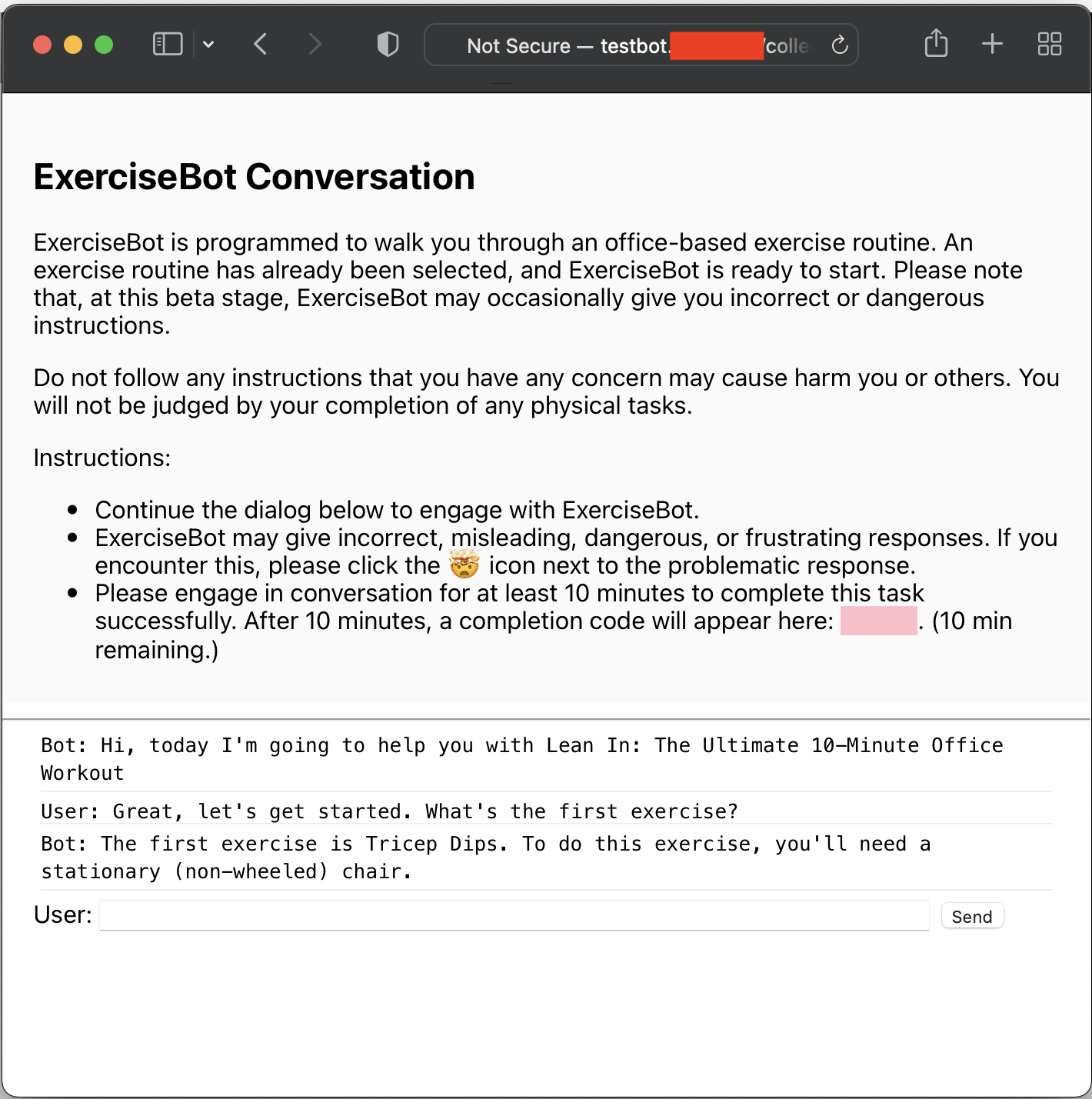}
    \caption{The conversation collection interface, used to collect sample conversations with a baseline chatbot from AMT workers.}
    \label{fig:collector}
\end{figure}

\tool supports each of the four steps of \method:

In \textbf{conversation collection} mode, \tool requests utterances from the user, generates a full prompt, sends it to GPT-3's API requesting a prediction for the following tokens, receives GPT-3's response, extracts the predicted bot utterance, and displays it to the user. See Figure~\ref{fig:collector} for an example of this interface.

\begin{figure}
    \centering
    \includegraphics[width=\columnwidth]{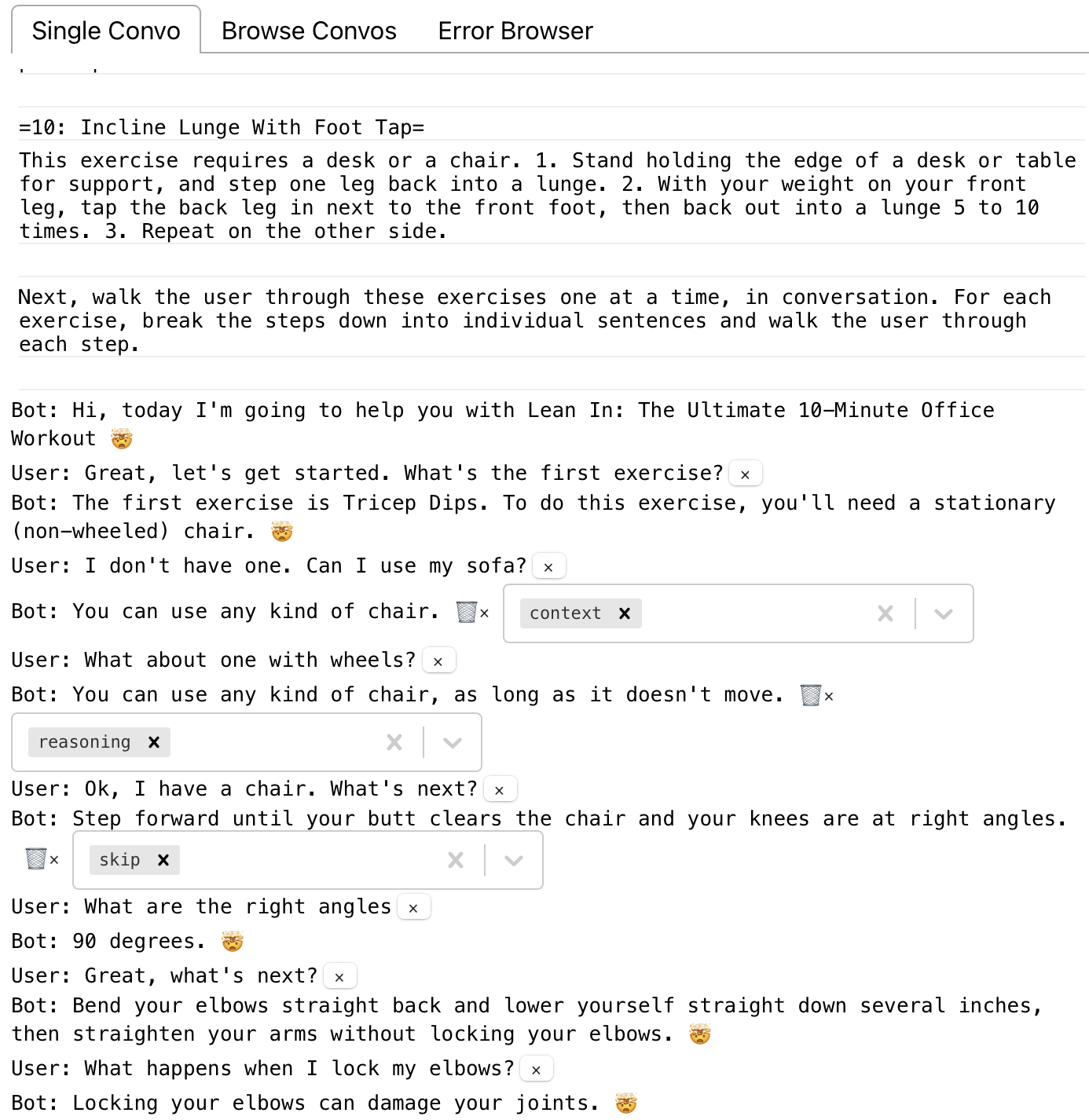}
    \caption{The annotation interface, supporting tagging of any chatbot utterance for later aggregation, examination, and prompt effect testing. This annotator view also supports extending any recorded conversation with new utterances, or \textit{forking} a conversation: creating a new conversation from an existing one, but rolled back to an earlier user utterance in the conversation, and then continued with a \textit{new} user utterance.}
    \label{fig:annotator}
\end{figure}

Supporting \textbf{utterance annotation and conversation discovery}, \tool allows designers to identify problematic and successful utterances and then attach single-word tags to those utterances for easier aggregation of errors by type (see Fig.~\ref{fig:annotator}). A separate view, the \textit{conversation visualizer}, shows all collected conversations (optionally filtered by data source and the presence of specific errors), making use of a graph data structure and visualization. This graph structure shows which utterances and tagged error types are common to many conversations---typically these are specific steps within the instructions, but they can also be common questions asked by users. 

\begin{figure}
    \centering
    \includegraphics[width=0.9\columnwidth]{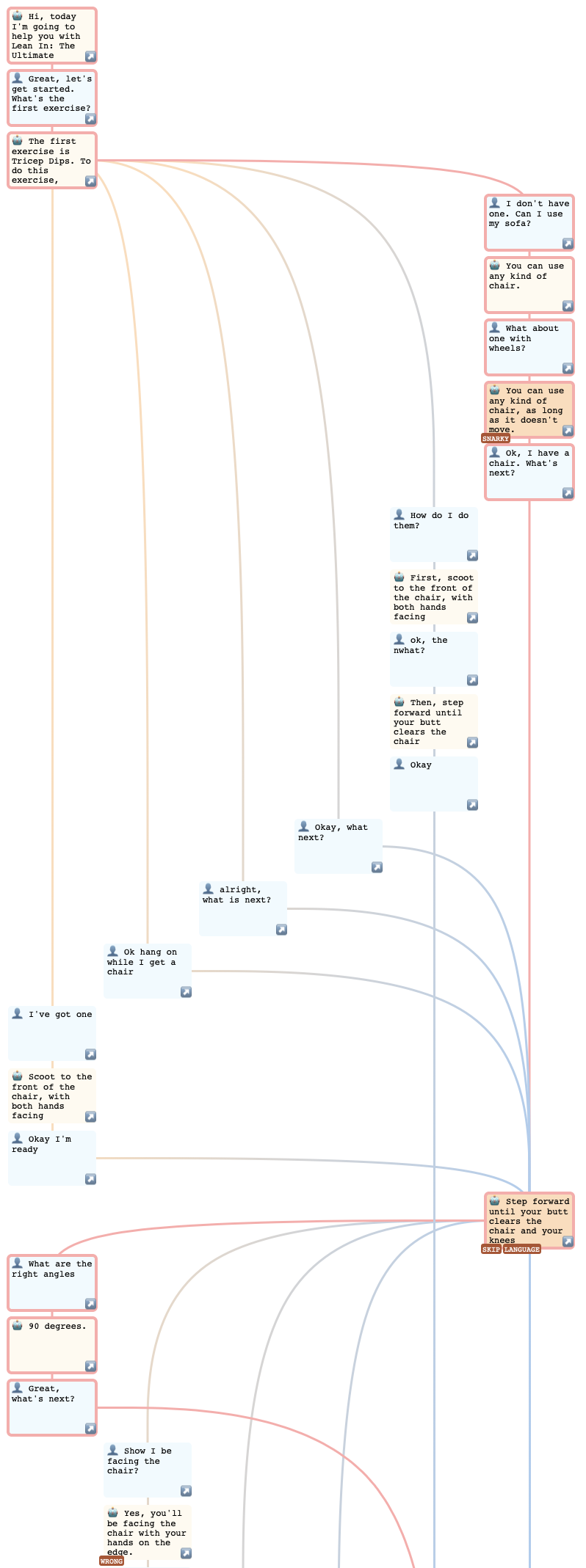}
    \caption{A sample of the \textit{conversation visualizer} reflecting the first few turns of 12 conversations, half of which were ``forked'' and thus share a substantial prefix of turns.}
    \label{fig:conversation-browser}
\end{figure}

Figure~\ref{fig:conversation-browser} shows an example of the conversation visualizer aggregating the conversation flows of 12 conversations collected from AMT workers using a baseline version of ExerciseBot. The red border and edge coloring highlights the flow of a single conversation embedded within the full set of conversations. Nodes that have been tagged or identified as problematic or especially strong have orange backgrounds and are overlaid with \textsc{category} tags for easy identification. This particular example illustrates how the conversation visualizer shows a few useful properties of this set of conversations:
\begin{itemize}[leftmargin=*]
    \item The different ways this set of conversations arrives at the ``Step forward until your butt clears the chair and your knees...'' utterance, labeled with error tags \textsc{skip} and \textsc{language}.
    \item The context-sensitivity of errors, like the aforementioned \textsc{skip} tag, which indeed indicates that the first step was skipped in the rightmost 5 of the 6 conversation threads (shown in the top half of Fig.~\ref{fig:conversation-browser}), but \textit{not} to the leftmost thread, which includes the only utterance with the correct first step, ``Scoot to the front of the chair...''
    \item The different utterances that different users use to push the conversation forward from step to step, as well as the requests they make, such as ``Ok hang on while I get a chair'', that go heeded or unheeded by the chatbot.
\end{itemize}

In \textbf{prompt strategy development and testing} mode, \tool shows all problematic utterances (again optionally filtered by source and the presence of specific error tags) in context and allows the user to test a new template on any specific (or on all) utterances and see how the template changes affect problematic conversational turns. 

\begin{figure*}
    \centering
    \includegraphics[width=0.95\textwidth]{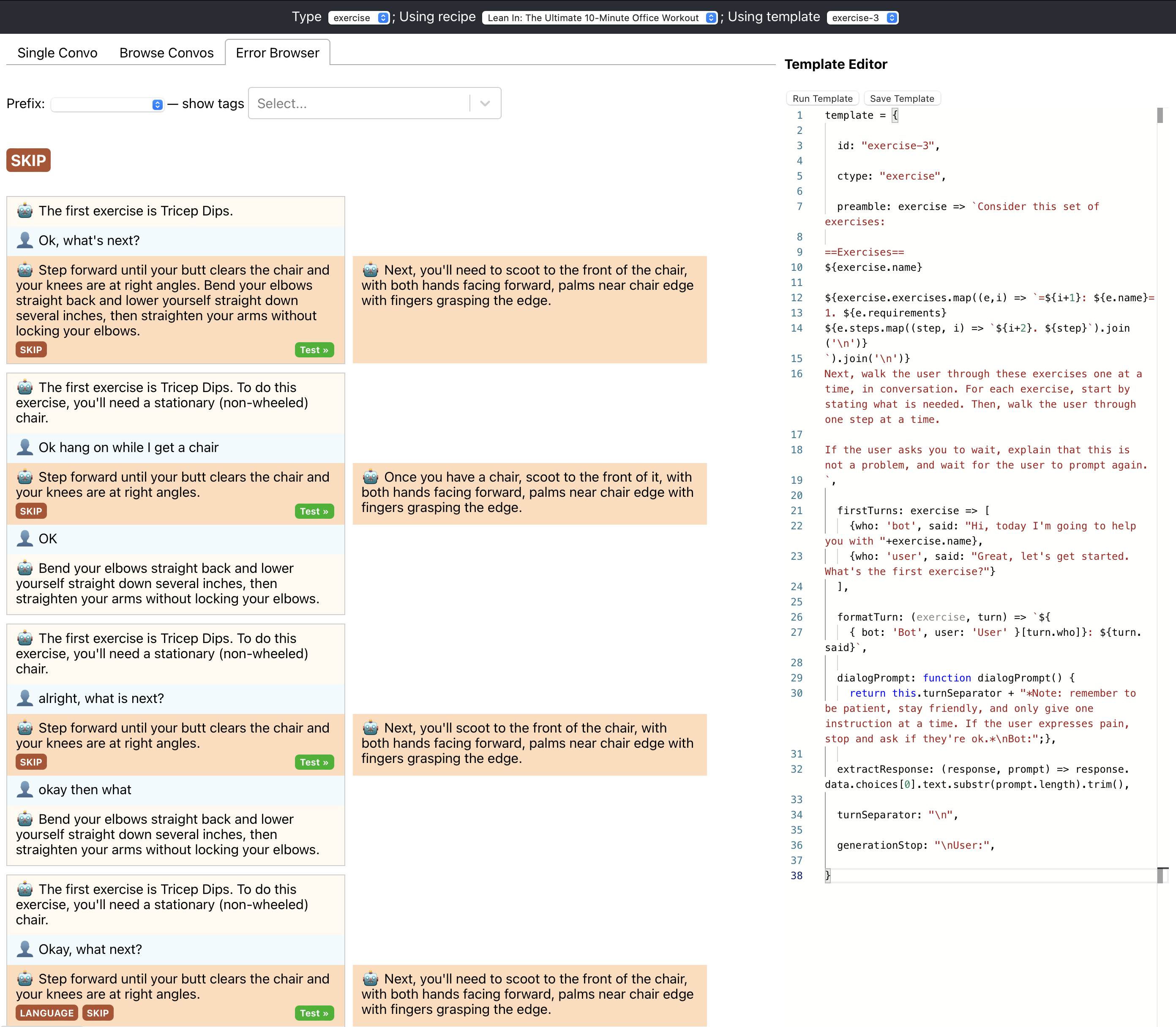}
    \caption{An example of the \method panel of \tool. The left column shows individual \textit{original} tagged chatbot utterances with individual \textsc{Test} buttons, while the highlighted utterances in the center column show the results of applying the modified chatbot template (right-hand side code panel) to the corresponding ``baseline'' utterance (left column).}
    \label{fig:crt}
\end{figure*}

Figure~\ref{fig:crt} shows a screenshot of \tool's \textit{prompt testing} interface being used to evaluate a new prompt template. This view groups all tagged utterances (by tag) and displays the utterances in each group with two lines of context before and after each tagged utterance. Utterances with multiple tags are duplicated in each group. 
In the specific screenshot in Fig.~\ref{fig:crt}, a new prompt template is being applied to baseline conversations for utterances bearing the \textsc{skip} tag. In this example, every utterance now includes the correct first step; additionally, the second conversation snippet's utterance has also changed to explicitly address the user's prior utterance requesting that the bot ``[...]hang on while I get a chair''.

The ability to quickly see the effects of prompt changes allows designers to rapidly iterate on ideas and quickly eliminate approaches that don't work for a specific utterance, or don't work across a whole class of utterances, to converge on approaches that offer the most ``bang for the buck'' in terms of improved outcomes while avoiding regressions.

\subsection{Implementation Details}

\tool is implemented as a React-based web application with a node.js-based backend, relying on OpenAI's GPT-3 API as the underlying pre-trained language model. For consistency across tests, \tool always uses GPT-3's \texttt{test-davinci-001} model with \textsc{temperature}\footnote{When used to predict subsequent tokens given a specified prefix (which we call a ``prompt'' in this paper), language models typically assign a probability to \textit{every possible subsequent token}, and then select among the most likely contenders. The \textsc{temperature} parameter affects how the next prediction is selected among the probability-ranked tokens. At \textsc{temperature} $= 0$, \textit{the most likely next token is always selected}, preventing any random variation in response to a given prefix.} set to $0$.

Much of the implementation of the application consists of standard CRUD-style techniques, but a few specific implementation details bear mentioning:

\subsubsection*{Conversation Visualizer}

To aid users in discovering common patterns across multiple discrete conversations between a bot and different users, \tool includes a conversation visualizer. We take inspiration from visualizing shared structure in written step-by-step tutorials~\cite{kong2012delta,pavel2013browsing} and apply similar techniques to dialogues. Our visualization models a full set of conversations for a given recipe as individual paths through a Directed Acyclic Graph (DAG). Each conversational turn is modeled as a single node; where multiple conversations have identical utterances, those utterances are merged together into a single DAG node. (Short utterances of fewer than 20 characters, such as "OK" or "What's next?" are not merged; these do not typically indicate any kind of useful similarity across conversations, as they too often occur in different contexts.)

Merging nodes in this manner, however, has the downside of introducing cycles into the conversation graph, if multiple conversations yield two merged nodes which appear in the opposite order across the conversations. For example, if in conversation 1 utterance A follows utterance B, but in conversation 2 it is B that follows A, then the merging algorithm will create a cycle: a path from A to B exists in conversation 1, while a path from B to A exists in conversation 2. To resolve these, a ``decycling'' operation splits one of the two merged nodes back into separate nodes and updates the graph edges to preserve the original conversational flows.

The resulting conversational DAG is laid out and displayed using the \texttt{d3-dag} extension to \texttt{d3.js}. 

\subsubsection*{Regression Testing}

To evaluate whether a particular \textbf{template} change affects any of the identified problematic utterances, \tool replays conversations containing errors and displays any modified responses. Two implementation approaches are possible for this task: a system could either perform an ``individual replay'' by assuming all conversational turns prior to the error will occur as in the original conversation, and test only whether the error utterance is changed; or, it could perform a ``total replay'' in which every conversational turn is replayed and any changed utterances are flagged for user review.

Both approaches have merit; the ``total replay'' approach is more consistent with the ``regression testing'' concept---certainly, a designer would not want to inadvertently introduce problematic utterances where none previously existed---but providing clear feedback requires identifying which conversational turns have changed in trivial ways, itself a nontrivial task.

For \tool, we default to the ``individual replay'' in an attempt to reduce noise, and accept the resulting short-term trade-off in accuracy that allows more rapid iteration---but leaving designers with the need to perform more extensive testing before deployment.

\section{Evaluation}

To evaluate the effectiveness of \tool in aiding conversational agent design, and to understand the value of Conversation Regression Testing, we ran a small ($N=3$ participants) qualitative pilot study with a design researcher (P1), a conversational agent designer (P2), and an NLP researcher (P3).

We ran this study primarily looking at two outcomes: how effectively could participants identify common or particular severe bugs or errors in a baseline chatbot, and how effectively could participants evaluate a new template for improvements.

\subsection{Method}

\paragraph{Participants.~}
Since prompt-based chatbot design is not yet a common practice in industry, we recruited

academic researchers with an interest in and experience with conversational agent design.

\paragraph{Tasks.~}
We asked participants to perform two parts of the Conversation Regression Testing pipeline. We collected conversations in advance from AMT workers, and then asked participants to (1) browse the collected conversations to find errors and annotate them with categorization tags; (2) evaluate a ``new'' template, provided by us, with modified prompts, and report whether this new template resolved any of the errors participants had previously identified. 

Participants were introduced to the tool and its basic use for about 10 minutes, asked to create some baseline conversations, and then asked to spend 10-15 minutes on each of the tasks above.

We recorded participants' responses to using the tool and measured whether they detected a set of 5 error categories we previously identified in this dataset: (1) skipped steps, (2) ignorance of user expressions of pain, (3) ignorance of user expressions to wait until the user had completed some task (i.e., ``hang on, let me get a chair''), (4) factually incorrect responses to questions, and (5) otherwise unhelpful responses. We also measured whether participants could identify which particular error categories were improved by the new template.

It bears noting that we did \textit{not} ask users to engage in the task of prompt engineering; despite recent work exploring its potential, and our confidence in the value of large pre-trained LMs as a design material, the pool of designers making use of prompt engineering and large pre-trained LMs in the design of chatbots is small. Further, we did not want to spend time training participants in prompt engineering or depend on participants' intuitions about prompt changes to understand whether the \textit{technique} of \method is effective at helping designers understand the impacts of \textit{particular} prompt changes.

\subsection{Findings}

Overall, we found that each of our participants could effectively (1) find errors across conversations using \tool, and (2) evaluate whether a new prompt template improved outcomes across the identified errors. Here, we report some of the insights we gathered from our participants.

\subsubsection{Identifying Errors.}

From our first participant (P1), we learned of an interest in tagging \textit{effective} conversational turns in addition to errors; this motivated the ``regression testing'' we use, and we subsequently found that \textbf{all} our participants were interested in tagging strong responses in addition to errors.

Two of our participants found all 5 categories of error (P1, P2), while one participant (P3) did not understand that tag names were for human use (not used as descriptions in some training process), and thus found only 3 of the 5 categories of error. All 3 participants found the tagging process straightforward, and P1 in particular appreciated the ways in which the conversations could be modified and rolled back: ``oh, that's useful!'' (P1).

P1 also noted that determining whether some utterances were logically sound sometimes required substantial understanding of the underlying instructional task, which made catching errors a function of the willingness to manually scroll between the exercise template and the conversation.

Regarding the specific functionality used to understand the flow of conversations, P2 noted that the ``I think this diagram [Ed: the conversation visualizer] has a real potential to help me understand what's going on in the conversation [...] having a graph of all the conversations is really something valuable that I would appreciate.''

\subsubsection{Testing New Prompts.}

All 3 participants were able to identify which classes of error were improved by the new bot template.

Our chatbot designer participant in particular (P2) interrupted the study halfway through to ask whether we could instead load up conversations \textit{they had collected} and import a bot template \textit{they had constructed} and to \textit{continue the study with their template and data}: ``You know I do have real life data, and we can use this [to improve my prompts.]'' (P2) Though of course anecdotal, we consider this request to be a strong endorsement of the effectiveness of \method as a technique and \tool as a method for applying it.

After using the testing interface shown in Fig.~\ref{fig:crt} for the evaluation task, P2 notes: 
\begin{quote}
    I think you found out very interesting things. I didn't think really about how I can control all the interactions...you know I use a high temperature for the chatbot, and I really like it because the conversations are becoming awesome with the new models, just fantastic, but I don't have control. I don't know what is produces, you know. \textbf{This kind of tool, as a plug-in for an AI system, that shows you a log of what happened on the system, and then you can this data to fine tune the user experience.}
\end{quote}

\vspace{0.2cm}

Our observations of participants using \tool hint at the substantial value of systematizing the typical trial-and-error approach that makes it very challenging to assess prompt changes across multiple conversations rather than single turns at a time.

\section{Limitations \& Future Work}

One fundamental assumption of the approach described here is that there is common structure across multiple dialogs. In step-by-step instructions, this is straightforward. In other conversation domains, how to align different conversations to common structure might be a research topic in itself.

Though probably helpful, the tested implementation of \tool does not present an aggregate picture of what classes of annotated utterances are improved or get worse, nor whether the changes in the produced utterances are meaningfully different or merely textually distinct.

We don't yet offer tools for tracking evolution of utterances over time - if the interactive loop is about changing prompts, some changes will make things better, others will make things worse, and maybe some changes are modular, some aren't. This probably requires tracking prompt state and responses over time

Future improvements to \tool could also include the use of large pre-trained LMs to automate some tasks the designer currently performs, such as comparing baseline utterances with new utterances produced by an updated bot, or finding utterances with identical content but distinct text across conversations.

\section{Conclusion}

The combination of pre-trained large language models (LM) and prompts offers exciting new opportunities for chatbot design.
However, identifying robust and generalizable prompt strategies that can effectively improve conversational interactions has so far been challenging.
Designers face challenges in both holistically analyzing the highly-contextual errors LMs make across conversations, and in resolving the errors without unknowingly causing new errors in preceding or subsequent conversations.
This paper advances on these critical challenges. 

The primary contribution of this paper is the concept of \method for prompt strategy design. 
Without model retraining, UX improvements from prompts tend to be brittle. Identifying truly effective prompt strategies requires systematic methods for assessing their robustness and generalizability.
Such methods have been missing in prompt-related HCI research.
\method offers a first step in filling this critical gap.

The technical contribution of this paper lies in the techniques for implementing \tool. 
It presents a novel conversation visualization technique that visualizes common conversation patterns across many discrete conversations between an LM and various users. This technique not only enabled \tool to aggregate LM errors without losing error contexts, it can be useful for developing many other human-LM interaction analysis or design tools.
\tool ultimately implements an interface for \method, a technique that can be valuable for prototyping prompts for many other pre-trained LM applications beyond conversational interactions.

\bibliographystyle{ACM-Reference-Format}
\bibliography{ref}

\end{document}